\documentclass[runningheads]{llncs}
\usepackage[T1]{fontenc} 
\usepackage{graphicx} 

\usepackage[disallowspaces]{mathtools}
\usepackage{amssymb} 
\usepackage{booktabs} 
\usepackage{forest}
\usepackage{longtable}

\newcommand{\limp}{\rightarrow}
\newcommand{\lequiv}{\leftrightarrow}

\newcommand{\false}{\mathit{false}}
\newcommand{\true}{\mathit{true}}
\newcommand{\expl}{\mathit{expl}}
\newcommand{\intr}{\mathit{intr}}
\newcommand{\rewrite}{\mathit{rewrite}}

\newcommand{\connectsto}{\mathit{connectsto}}
\newcommand{\leftbranch}{\mathit{leftbranch}}
\newcommand{\rightbranch}{\mathit{rightbranch}}
\newcommand{\partof}{\mathit{partof}}
\newcommand{\before}{\mathit{before}}
\newcommand{\ahead}{\mathit{ahead}}
\newcommand{\inrear}{\mathit{inrear}}
\newcommand{\occupied}{\mathit{occupied}}
\newcommand{\proceed}{\mathit{proceed}}
\newcommand{\routelocked}{\mathit{routelocked}}
\newcommand{\conflict}{\mathit{conflict}}
\newcommand{\entry}{\mathit{entry}}
\newcommand{\first}{\mathit{first}}
\newcommand{\inconflict}{\mathit{inconflict}}
\newcommand{\pointsok}{\mathit{pointsok}}
\newcommand{\safe}{\mathit{safe}}
\newcommand{\unoccupied}{\mathit{unoccupied}}

\newcommand{\pt}[1]{\mathit{pt#1}}
\newcommand{\tc}[1]{\mathit{tc#1}}
\newcommand{\rt}[1]{\mathit{rt#1}}
\newcommand{\si}[1]{\mathit{si#1}}
\newcommand{\green}[1]{\mathit{green#1}}
\newcommand{\red}[1]{\mathit{red#1}}
\newcommand{\lock}[1]{\mathit{lock#1}}
\newcommand{\track}[1]{\mathit{track#1}}
\newcommand{\ptleft}[1]{\mathit{left#1}}
\newcommand{\ptright}[1]{\mathit{right#1}}

\newcommand{\blue}{\mathit{blue}}
\newcommand{\neighbours}{\mathit{neighbours}}

\begin{document}

\title{Why does it fail? \\Explanation of verification failures 
}
\titlerunning{Why does it fail ...}
\author{Lars-Henrik Eriksson
}
\authorrunning{L.-H. Eriksson}
\institute{Department of Information Technology, Uppsala University, Uppsala, Sweden
\email{lhe@it.uu.se}}
\maketitle

\begin{abstract}
Satisfiability solving is a common technique for formal verification forming the basis of many proof and model checking systems. Failure to show a proof obligation will produce a counterexample or failure trace with typically many thousands or even millions of boolean variables. Interpreting such a counterexample poses a challenge. Even if the individual variables are all understood, it is difficult to form a ``big picture'' of the situation causing the failure. We consider the case where verification conditions are expressed using concepts from a formal application domain model in a language based on
predicate logic or a similar language. We introduce a method to explain verification failures in application domain terms. A measure of the relative relevance of predicates is used to extract the parts of a formula most likely to contribute meaningfully to the explanation.  Dependencies between predicates are used to form a branching sequence of successive explanations. These explanations can help a practitioner find  faults in the system being verified. The method is demonstrated on examples and compared to other methods.
\end{abstract}

\section{Introduction}

Satisfiability (SAT) solving is a fundamental proof technique that is  used as the underlying method of many proof and model checking tools used for formal modelling and verification, notably the Alloy Analyzer \cite{DBLP:journals/cacm/Jackson19}. Many verification problems expressed in a rich specification language can be naturally translated into propositional logic. Even though such translations typically give rise to very large SAT problems with hundreds of thousands or even several million propositional variables, modern SAT solving engines solve such problems efficiently \cite{SAT2025}. When formal verification fails, the tool will provide a counterexample – a set of facts, such as combination of inputs and outputs or parts of a trace, which disagrees with the formal specification.

When traditional testing fails, the specific situation that leads to an error is given by the test case. In formal verification there are no test cases as such, instead a particular situation where the error occurs is implicit in the counterexample. It is then necessary to analyse the counterexample to reconstruct a ``test case'' after the fact to be able to understand why the failure occurred. Given the large amount of low-level data involved in the counterexample and that most of it will be unrelated to the fault, this is a daunting task to take on.

A \textit{formal domain model} \cite{bjorner2023domain} is a mathematical description of the environment (application domain) in which a computer system operates. A well-structured model helps the software engineer to better understand the application domain and thus to write better formal requirements specifications of a software system. It also makes it easier for application domain experts who are not themselves formal methods practitioners to understand formal specifications.

A formal domain model can describe the environment on different levels. For example, in the \textit{railway domain}, a domain model can describe the kinds of entities and their possible relationships such as the physical tracks, signalling systems, train movements, time tables, rolling stock management, reservations, ticketing etc. depending on the purpose of the software to be designed \cite{penicka2006towards}. Combining the domain model with data about \textit{concrete} entities and relations, e.g. descriptions of the actual tracks, signals etc. of a particular railway \cite{domains,GTO}, yields a model of the environment in which a specific computer system operates.

Formal domain models can be used to create high-level descriptions of counterexamples. Low-level counterexample data can be translated back to concepts defined in the domain models. For example, in the railway domain individual signal lamp status signals can be translated back to abstract signal aspects. However, there need also be a mechanism to filter out relevant data from the counterexamples. As an example, if a railway signalling system omits a necessary check of a status signal, causing a high-level requirement to be violated, an explanation of the counterexample could point to that particular status signal.

We have developed a method to create such explanations. Compared to other methods, the main innovation is the use of a concept of ``interest'' to extract the information most likely to give a useful explanation. The method has been implemented and experimentally tested on the verification tool GTO \cite{GTO,saadoun2022implementation}.

The method can of course also explain ``positive'' results e.g. why a formula is true, but it was developed primarily with failure explanations in mind.

This article is organised as follows: In section \ref{foundations} we summarise our logical formalism. Section \ref{finding} describes our method for explanation-finding while section \ref{examples} given an applied example. Section \ref{discussion} discusses the method and compares it to an alternative method. Section \ref{relatedwork} discusses relevant existing work. Finally, in section \ref{conclusions} we give conclusions and directions for further work.

\section{Foundations}
\label{foundations}

We consider the problem of finding explanations to failed formal verification problems in many-sorted first-order (predicate) logic with finite universes (bounded quantification), such as can be found using SAT solvers. In this section we summarise the necessary logical concepts. They are for the most part standard and are included as a reference to the particular variant of formalism we use.

The \textit{formulæ} of our logic are constructed in the standard way from a set of predicate symbols including equality, constant symbols, variables, and the operators $\lnot, \land, \lor, \limp, \lequiv, \forall, \exists$. Predicate symbols (or simply ``predicates'') take zero or more variables or constants as arguments. (Zero argument predicates are the same as propositional symbols.) For simplicity, we do not include function symbols in the language, but generalisation to a logic with function symbols is straightforward. An \textit{atom} is a predicate applied to a set of arguments (constants or variables). A \textit{literal} is an atom or a negated atom. We will write predicates, variables and constants in lowercase and use uppercase names for formulæ.

Our logic also has \textit{definitions} of the form $p(x_1,\dots,x_n) \equiv P, n\geq 0$ where the $x_i$ are distinct variables. Logically, they represent the formulæ $$\forall x_1,\dots,x_n.(p(x_1,\dots,x_n)\lequiv P)$$ but provide additional information to our explanation algorithm. 

\textit{Free and bound variables} are defined in the usual way. Variables range over objects of some particular subset of a \textit{universe}, depending on their \textit{sorts}. We assume that every object in the universe is associated with an individual constant symbol. To simplify the presentation, we will identify objects with their corresponding constants.  A \textit{closed formula} is a formula with no free variables.

We will sometimes use sets of formulæ in place of a single formula. Such a set should be understood as a representation of the conjunction of all formulæ in the set. The empty set represents a formula that is always true.

An \textit{interpretation}, $\mathcal{I}$, is an assignment of truth values to ground (variable-free) atoms. Thus under an interpretation any closed formula has a particular truth value. In the sequel, we will identify an interpretation with the set of atoms true under the interpretation. The standard definition of an interpretation in predicate logic is a structure which includes the universe. Here we separate the two and take the universe (and the sorts) as given. This is motivated below.  

An interpretation $\mathcal{I}$ \textit{satisfies} a closed formula $P$, formally $\mathcal{I}\vDash P$, iff $P$ is true under $\mathcal{I}$. 
A formula is \textit{satisfiable} if there is a satisfying interpretation, otherwise \textit{unsatisfiable}.
Note that if $\mathcal{I}\not\vDash P$, then $\mathcal{I}\vDash \neg P$ and vice versa. 

In the literature, the term \textit{model} is used with two different meanings. One is for a set of logical axioms -- a \textit{theory} -- that describes some system, the other is for an interpretation such that all the axioms of a theory are true. Here we will only use the term ``model'' in the first sense. When the second sense is needed, we will talk about a ``satisfying interpretation''.

$R$ is a \textit{logical consequence} of $P$ iff every interpretation that satisfies $P$ also satisfies $R$. Formally, we write $P\vDash R$. As a special case, $\vDash R$ means that $R$ is true in every interpretation or, equivalently, that there is no interpretation that satisfies  $\neg R$. $P\vDash R$ means that every interpretation either makes $P$ false or $R$ true. This is the same as $\vDash P\limp R$, so to show $P\vDash R$, it suffices to show that there is no interpretation satisfying  $ \neg(P\limp R)$ or, equivalently,  $P\land\lnot R$.

SAT solvers find satisfying interpretations of formulæ in propositional logic. To use a SAT solver with predicate logic, formulæ must first be transformed into a propositional formulæ. With a finite universe, this can be done by rewriting every quantified (sub)formula as a conjunction (for $\forall$) or a disjunction (for $\exists$) of the set of instances of the quantified formula obtained by substituting bound variables with each constant of the universe belonging to the sort of the variable. Formally, $\forall x.P$ is rewritten as $P[c_1/x]\land\dots\land P[c_n/x]$ where $\{c_1,\dots,c_n\}$ is the set of constants belonging to the sort, and analogously for $\exists$.

With all quantified formulæ rewritten, the atoms are ground and can be regarded as a propositional symbol so the rewritten formula can be used by a SAT solver. If the formula is satisfiable the SAT solver will give a truth value to each atom, i.e. an interpretation. The fact that the rewriting uses a particular universe motivates the separation of the universe from the interpretation.

A formal verification problem involves showing that a satisfying interpretation of a model of the system to be verified, $P$, meets requirements expressed by a verification condition $R$, given an application domain model $D$. Some predicates are used to represent inputs to the system, others are used to represent outputs or internal states. Thus an interpretation expresses a certain combination of inputs, outputs and states. A system is correct if every input-output-state combination made possible by the system is also allowed by the requirement. Formally, we must show that every satisfying interpretation of the system and the domain also satisfies the requirement, or $P,D\vDash R$. This is the same as $\not\vDash P\land D\land\lnot R$. If there actually is a satisfying interpretation of $P\land D\land\lnot R$, then it describes a situation occurring in the system which is not allowed by the requirements. Our goal is to describe ways to explain why that situation has arisen.

\section{Finding explanations}
\label{finding}
Explanations are always in the context of a particular interpretation. When stating that a formula is true or false, it is with respect to that interpretation. 

Our method proceeds in two alternating phases. One phase gives an explanation for a particular formula, while the other phase creates a tree of successive explanations. We will introduce the first phase by an example.

\begin{example} \label{motivatinginterest}
Let $p$ be the characteristic predicate of a set and let $q$ represent an interesting property. Assume a single sort with universe $\{a, b, c\} $ and an interpretation such that $p(a), q(a), q(c)$ are all true and $p(b), p(c), q(b)$ are all false. That is, the set characterised by $p$ is $\{a\}$ and $a$ and $c$ both have the property.

Now, the formula $\forall x.(p(x) \limp q(x))$ is true under the interpretation. What atom is intuitively the most important one for understanding why the formula is true? Arguably, it is $q(a)$. Why? Atoms of the form $p(x)$ serve as ``filters'' to decide for what objects we should check the property and are thus less interesting in themselves. As $b$ and $c$ are ``filtered out'', $q(b)$ and $q(c)$ do not affect the truth value of the formula and are thus also less interesting. We say that the set $\{q(a)\}$ is the explanation of why $\forall x.(p(x) \limp q(x))$ is true (This example will be described in more detail after the definition of the algorithm.)
\end{example}

Our algorithm will compute exactly that explanation. It works by selecting ``interesting'' parts of the formula using  two criteria. One criterion is that an immediate subformula is interesting if changing its truth value would change the truth value of the whole formula.

\begin{example}
Suppose that $P$ is true and $Q$ is false. Then the explanation of $Q$ is also an explanation of why $P\land Q$ is false, as changing the truth value of $Q$ to true will also change the truth value of $P\land Q$ from false to true.
\end{example}

If changing the truth value of \textit{any} single subformula will change the truth value of the whole formula, then instead an ``interest measure'', $\intr$, on formulæ is used. The explanation will be taken from the highest interest subformula. If both subformulæ have the same measure, then the explanations of the subformulæ together is the explanation of the whole formula.

\begin{example}
Suppose that $P$ and $Q$ are both true and that $\intr(P)=\intr(Q)$. Then the explanations of $P$ and $Q$ together is the explanation of why $P\land Q$ is true. Suppose instead that $\intr(P)>\intr(Q)$. Then only the explanation of $P$ is an explanation of why $P\land Q$ is true.
\end{example}

If the truth value of a formula is unchanged unless the truth value of \textit{both} subformulæ are changed, then instead we use the explanation of the subformula with the \textit{least} interest measure. If both subformulæ have the same measure, one explanation is chosen arbitrarily.  This is motivated later in this section and further discussed in section \ref{discussion}.

\begin{example} \label{truetrueor}
Suppose that $P$ and $Q$ are both false and that $\intr(P)=\intr(Q)$. Neither $P$ nor $Q$ provides an explanation to why $P\land Q$ is false according to our first criterion as changing the truth value of either (alone) does not change the truth value of $P\land Q$. As the falsity of $P$ and $Q$ separately explains the falsity $P\land Q$, the explanation of one of them is arbitrarily chosen. Suppose instead that $\intr(P)>\intr(Q)$. Then the explanation of the subformula with the \textit{lesser} interest, $Q$, is chosen as the explanation of why $P\land Q$ is false.
\end{example}

These principles are applied recursively in a bottom-up manner. Initial interest measures are applied to predicates and explanation formulæ inherit the measure of their explanation subformulæ. Atoms are negated in explanations if they are false. Formally, the explanation function $\expl$ and the interest measure function $\intr$ are defined together as:

\begin{definition}[$\expl$ and $\intr$]
$$
\begin{array}{l}
\left.
\begin{array}{l}
\expl(A) = 
\begin{cases}
\{A\} & \textup{ when $A$ is true} \\
 \{\lnot A\} & \textup{ when $A$ is false} 
 \end{cases} \\
\intr(A) = \textup{(depending on the application)}
\end{array}
\right\}
\textup{if $A$ is an atom}
\\
\left.
\begin{array}{l}
\expl(\lnot P) = \expl(P) \\
\intr(\lnot P) = \intr(P)
\end{array}
\right\} 
\\
\left.
\begin{array}{l}
\expl(P \land Q) =  \\
\intr(P \land Q) =
\end{array}
\right\}
\textup{according to Table \ref{expltable}}
 \\
\left.
\begin{array}{l}
\expl(P) = \expl(\rewrite(P)) \\
\intr(P) = \intr(\rewrite(P))
\end{array}
\right\}
\textup{in other cases, see definition \ref{rewrite}}
\end{array}
$$
\end{definition}

\noindent
The auxiliary function $\rewrite$ is defined as:

\begin{definition}[$\rewrite$] \label{rewrite}
$$
\begin{array}{ll}
\rewrite(P_1 \lor \dots \lor P_n) = \lnot (\lnot P_1 \land \dots \land \lnot P_n) \\
\rewrite(P \limp Q) = \lnot (P \land \lnot Q)\\
\rewrite(P \lequiv Q) = (P \limp Q) \land (Q \limp P) \\
\rewrite(\forall x.P) = P[c_1/x] \land \dots \land P[c_n/x] \\
\rewrite(\exists x.P) = \lnot (\forall x.\lnot P)  \\
\textup{where the $c_i$ are the constants belonging to the sort of $x$.}
\end{array}
$$
\end{definition}

\begin{table}[htp]
$$
\begin{array}{|c|c|c|c|c|}
\toprule
P & Q & \intr(P)\text{ vs. }\intr(Q)  & \expl(P\land Q) & \intr(P\land Q) \\
\midrule
\false & \false & {<} & \expl(P) & \intr(P) \\
\false & false & {=} & \expl(P)\text{ alt. }\expl(Q) & \intr(P) \\
\false & \false & {>} & \expl(Q) & \intr(Q) \\
\false & \true &  & \expl(P) & \intr(P) \\
\true & \false &  & \expl(Q) & \intr(Q) \\
\true & \true & {<} & \expl(Q) & \intr(Q) \\
\true & \true & {=} & \expl(P)\cup \expl(Q) & \intr(Q) \\
\true & \true & {>} & \expl(P) & \intr(P) \\
\bottomrule
\end{array}
$$
\caption{Explanations of ${\land}$} 
\label{expltable}
\end{table}

It may seem counterintuitive that in the false-false case for ${\land}$ in Table \ref{expltable}, we chose the explanation of the subformula with the \textit{least} interest measure. To see why, let's examine Example \ref{motivatinginterest} in more detail. According to the definition of $\rewrite$, the quantified formula $\forall x.(p(x) \limp q(x))$ is rewritten as a conjunction of all instantiations of the quantified subformula. Also, the implications are rewritten as negated conjunctions. We get a conjunction of three (true) conjuncts.
$$
\begin{array}{lll}
\lnot (p(a)\land \lnot q(a)) &~~& (1)\\
\lnot (p(b)\land \lnot q(b)) &~~& (2)\\
\lnot (p(c)\land \lnot q(c)) &~~& (3)\\
\end{array}
$$
The explanations of (1) and (2) are $\{q(a)\}$ and $\{\lnot p(b)\}$, respectively. As  (1) and (2) are both true and $p$ is less interesting than $q$, the explanation of their conjunction, $\lnot (p(a)\land \lnot q(a))\land\lnot (p(b)\land \lnot q(b))$ will be $\{q(a)\}$. But what about (3)? Both $p(c)$ and $\lnot q(c)$ are false, so there are two possibilities. If we use the most interesting one, $\lnot q(c)$, for the explanation of (3), the explanation of the conjunction of (1), (2), and (3) -- and thus the quantified formula --  will be $\{q(a), q(c)\}$ -- as $q(a)$ and $q(c)$ are of equal interest. That is not what we intended. If we (according to Table \ref{expltable}) instead choose the less interesting one, $\lnot p(c)$, from (3) it will be discarded when compared to $q(a)$, as both are true. The explanation of the whole formula will then be as intended, $\{q(a)\}$. 

In the second phase, we attempt to find further explanations of any defined explanation literals. For example, if $q$ is defined by $q(x) \equiv r(x) \lor s(x)$ the definition is used on $q(a)$ to obtain $r(a)\lor s(a)$. The explanation process is then repeated on this new formula. In Example \ref{motivatinginterest}, the explanation only included one literal, but generally there can be more than one. In that case we will get a branching structure of explanations -- an ``explanation tree''. The larger example in section \ref{examples} includes an explanation tree.

We have not attempted to give a formal characterisation of explanations. Ultimately, it is the practical usefulness of explanations that matters and not if they have a formal characterisation.
 
\section{A worked example}
\label{examples}

\begin{figure}[!htp]
\includegraphics{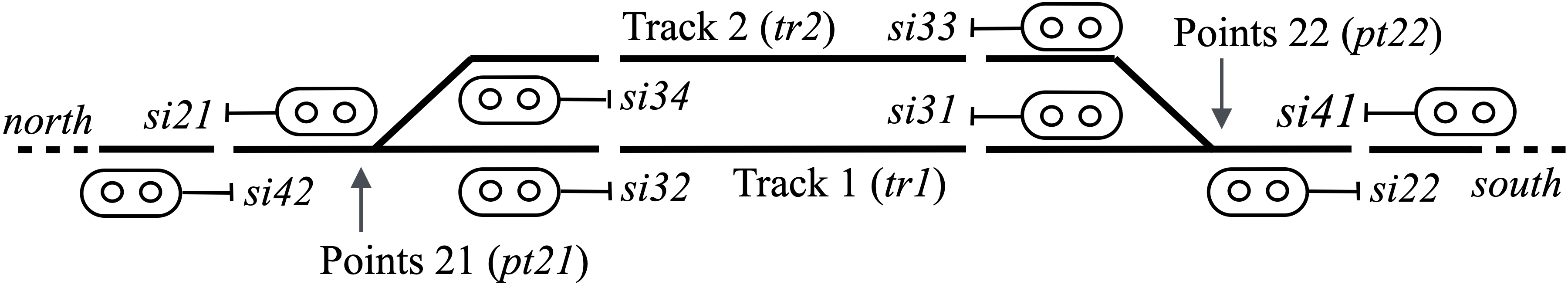}
\caption{Railway system}{}
\label{railway}
\end{figure}

As an application of explanation finding, we will use a fragment of a formal verification problem taken from the railway domain. The problem is to verify that the system controlling signals and points (switches) of the railway -- the \textit{interlocking} --  sets a signal to display a ``proceed'' aspect (green light) only when it is safe to do so. The methodology used is described in more detail in \cite{domains,GTO}. Consider the small railway system in Figure \ref{railway}.

The railway system is divided into \textit{units} (sections) while \textit{signals} control train movements. There are  \textit{train routes} -- sets of units over which trains move. Train routes start at a signal and extend to the next signal. They are identified by these signals, e.g. $\rt{2131}$ will designate the train route starting at signal 21 and ending at signal 31. In all, there are 8 different train routes in this example\footnote{Routes from signals 41 and 42 are not considered as they extend outside the railway system under consideration.}.

The universe has constant symbols for the designators of track units, signals and train routes. In the sequel, variables $u$ have the sort of track units, $p$ of points (a subset of track units), $s$ of signals and $r$ of train routes.

The verification condition is expressed in a generic way independent of a particular track layout using concepts from a formal domain theory. By supplying a description of the railway system, the verification condition becomes applicable to that particular system. Finally, the abstract model is combined with a model of the actual interlocking giving a complete verification problem. 

The domain theory has a set of predicates to describe the track layout. 

\begin{longtable}{p{3cm}lp{8.0cm}}
$\connectsto (u_1,u_2)$ && Track units $u_1$ and $u_2$ are adjacent to each other. \\
$\leftbranch(p, u)$ && In the left position,  points $p$ lead to unit $u$. \\
$\rightbranch(p, u)$ && In the right position,  points $p$ lead to unit $u$. \\
$\partof(u, r)$ && Unit $u$ is part of the train route $r$. \\
$\before(u, r)$ && A train enters train route $r$ from unit $u$. \\
$\ahead(s, u)$ && Signal $s$ is located at one end of  unit $u$, facing it. \\
$\inrear(s, u)$ && Signal $s$ is located at one end of  unit $u$, facing away from the unit.
\end{longtable}

The theory includes axioms that constrain these predicates to what is physically possible, e.g. that $\connectsto $ is symmetric. When using the domain theory in a particular application, they are defined to reflect the layout of the track system. 
For the particular railway system of Figure \ref{railway}, one definition would begin
$$
\begin{array}{l}
\partof(u,r) \equiv r=\rt{2131} \land (u=\pt{21} \lor u=\track{1})\ \lor \\
\phantom{\partof(u,r) \equiv\ }r=\rt{2133} \land (u=\pt{21} \lor u=\track{2}) \lor \dots\\
\end{array}
$$
It should be clear how the other predicates would be defined for this example.

Additional predicates are used for high-level representation of (abstract) input signals coming from the track system. A predicate instance is true if the corresponding input signal is present.

\begin{longtable}{p{3cm}lp{8.0cm}}
$\ptleft{}(p)$ && Points $p$ are in the left position. \\
$\occupied(u)$ && (Part of) a train is occupying unit $u$. \\
$\proceed(s)$ && Signal $s$ is showing a ``proceed'' aspect. \\
$\ptright{}(p)$ && Points $p$ are in the right position. \\
$\routelocked(r)$ && Route $r$ is reserved for a train movement. \\
\end{longtable}

Predicates representing various concepts can now be defined. 

\begin{longtable}{p{3cm}lp{8.0cm}}
$\conflict(r_1,r_2)$ &&Routes $r_1$ and $r_2$ can not safely be used by different trains at the same time. \\
$\entry(s,r)$ && Signal $s$ controls train entry into route $r$. \\
$\first(u,r)$ && $u$ is the first track unit of route $r$. (In the direction of train movement.) \\
$\inconflict(r)$ && Route $r$ is currently in conflict with another locked (reserved) route. \\
$\pointsok(r)$ && All points in route $r$ are set in the positions appropriate to the route. \\
$\safe(r)$ && Route $r$ is safe for train movements. \\
$\unoccupied(r)$ && The whole of route $r$ is free of trains. \\
\end{longtable}

Signals beginning a route show a ``proceed'' aspect (green light) into the route only when it is safe to do so. Entering a train route is considered safe whenever 
\begin{itemize}
\item the route is ``locked'', i.e. it is reserved for a train movement ($\routelocked$).
\item points in the route are in the correct position for train movement ($\pointsok$).
\item there are no other trains occupying any part of the route ($\unoccupied$).
\item no other locked train route is overlapping this route (negation of $\inconflict$).
\end{itemize}

\noindent The definitions of these predicates are
$$
\begin{array}{l}
\conflict(r_1,r_2) \equiv r_1\ne r_2 \land \exists u.(\partof(u,r_1)\land \partof(u,r_2)) \\
\entry(s,r) \equiv \\
~~\exists u.(\ahead(s,u)\land \before(u,r))\land \exists u.(inrear(s,u)\land \first(u,r)) \\
\first(u,r) \equiv \partof(u,r)\land \forall u_1.(\before(u_1,r)\limp \connectsto (u,u_1)) \\
\inconflict(r) \equiv \exists r_1.(r_1\ne r \land \routelocked(r_1)\land \conflict(r,r_1)) \\
\pointsok(r) \equiv \forall p.(\partof(p,r)\limp \\
~~(\exists u.(\leftbranch(p,u)\land (\partof(u,r)\lor \before(u,r)))\limp \ptleft{}(p))~~\land \\
~~(\exists u.(\rightbranch(p,u)\land (\partof(u,r)\lor \before(u,r)))\limp \ptright{}(p))) \\
\safe(r) \equiv \routelocked(r)\land \pointsok(r)\land \unoccupied(r)\land \lnot \inconflict(r) \\
\unoccupied(r) \equiv \forall u.(\partof(u,r)\limp \lnot \occupied(u)) \\
\end{array}
$$

\noindent Using these definitions, the verification condition can be stated as $$\forall s.(\proceed(s) \limp  \exists r.(\entry(s, r) \land \safe(r)))$$ I.e. if a signal shows a ``proceed'' aspect, then it must lead to a safe route.
 
We now turn to the interlocking to be verified. Suppose that an interlocking of the rail system in the figure correctly implements the verification condition and that the part controlling signal 21 is modelled by these definitions:
$$
\begin{array}{l}
\green{21} \equiv \lock{21} \land \lnot \lock{32} \land \lnot \lock{34} \land \tc{21} \land \\
\phantom{\green{21} \equiv} (\ptleft{21}  \land \tc{2} \land (\lnot \lock{22} \lor \ptleft{22}) \lor \\
\phantom{\green{21} \equiv (} \ptright{21} \land \tc{1} \land (\lnot \lock{22} \lor \ptright{22})) \\
\red{21} \equiv \lnot \green{21} \\
\end{array}
$$

The propositional symbols $\green{21}$ and $\red{21}$ represent outputs from the interlocking to the lights in the signal. The other propositional symbols represent states or inputs of the interlocking, being true when a route from a signal is locked (e.g. $\lock{21}$), when a set of points is in a particular position (e.g. $\ptleft{21}$) or when a track unit is free from trains (e.g. $\tc{21}$ is true if points 21 are free). 

Additionally, we need the axiom $$\lnot (\ptleft{21} \land \ptright{21})$$ to express the physical fact that points 21 can't be in both the left and the right positions simultaneously. (However, it is possible for them to be in an intermediate position.) 

Finally, we need to connect the predicates of the domain model representing ``abstract'' status with the concrete status signals of the interlocking.
$$
\begin{array}{l}
\ptleft{}(p) \equiv p=\pt{21} \land \ptleft{21} \lor p=\pt{22} \land \ptleft{22} \\
\occupied(u) \equiv u=north \land \lnot \tc{n} \lor u=\pt{21} \land \lnot \tc{21} \lor u=\track{1} \land \lnot \tc{1} \lor \dots \\
\proceed(s) \equiv s=\si{21} \land \green{21} \land \lnot \red{21} \lor  s=\si{22} \land green22 \land \lnot \red{21} \lor \dots \\
\ptright{}(p) \equiv p=\pt{21} \land \ptright{21} \lor p=\pt{22} \land \ptright{22} \\
\routelocked(r) \equiv \lock{21} \land (r=\rt{2131} \land \ptright{21} \lor r=\rt{2133} \land \ptleft{21}) \lor \dots \\
\end{array}
$$
The verification condition $\forall s.(\proceed(s) \limp  \exists r.(\entry(s, r) \land \safe(r)))$ is provable given these definitions and axioms. Now suppose we create a fault in the interlocking by dropping $\tc{1}$ from the definition of $\green{21}$. The verification condition will now be falsifiable and a SAT solver will find an interpretation where $\green{21}$ and $\ptright{21}$ are true while $\tc{1}$ is false.

Before applying the explanation algorithm on this counterexample, we have to decide on the interest measure of the different predicates. We use a ``syntactic'' approach to make the explanation process more automated and not require the user to manually assign interest values. Free predicates (predicates without a definition) are considered most interesting. In this example free predicates represent input signals to the interlocking -- directly pointing to a condition in the actual railway system. Predicates are then assigned lower interest measures according to how their definitions depend on other predicates.
Interest measures are assigned in descending order as follows:
\begin{itemize}
\item predicates without a definition (most interesting), e.g. $\tc{1}$ 
\item defined predicates that -- directly or indirectly -- depend on at least one predicate not having a definition, e.g. $\occupied$
\item defined predicates that  -- directly or indirectly -- depend on some other predicate, but only on predicates having a definition, e.g. $\conflict$
\item defined predicates that do not depend on other predicates, e.g. $\partof$
\item equality (least interesting)
\end{itemize}
               
To understand the reason for the failure of the verification condition, we first apply the explanation function $\expl$ to it, which gives
$$\{\proceed(\si{21}), \lnot \safe(\rt{2131}), \lnot \safe(\rt{2133})\}$$ Each literal starts a new branch in the explanation tree. Secondly, we use the definitions of $\proceed$ and $\safe$ on these literals to obtain the three formulæ

$$
\begin{array}{lll}
\si{21}=\si{21} \land \green{21} \land \lnot \red{21} \lor \\
~~\si{21}=\si{22} \land green(\si{22}) \land \lnot red(\si{22}) \lor \dots &&(1)\\
\lnot (\routelocked(\rt{2131})\land \pointsok(\rt{2131})\land \\
\phantom{\lnot (} \unoccupied(\rt{2131})\land \lnot \inconflict(\rt{2131})) &&(2) \\
\lnot (\routelocked(\rt{2133})\land \pointsok(\rt{2133})\land \\
\phantom{\lnot (}  \unoccupied(\rt{2133})\land \lnot \inconflict(\rt{2133})) &&(3) \\
\end{array}
$$

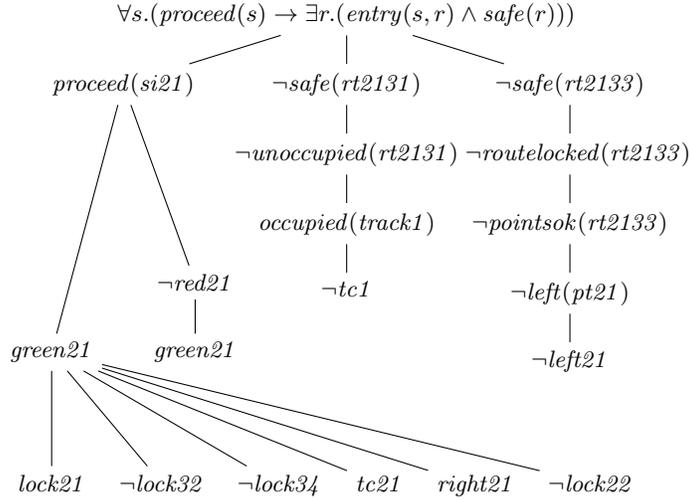
\begin{figure}[!htp]
$$
\begin{forest}
   [${\forall s.(\proceed(s) \limp  \exists r.(\entry(s, r) \land \safe(r)))}$
     [$\proceed(\si{21})$
       [$\green{21}$, l*=4, calign=first
         [$\lock{21}$, l*=2]
         [$\lnot \lock{32}$, l*=2]
         [$\lnot \lock{34}$, l*=2]
         [$\tc{21}$, l*=2]
         [$\ptright{21}$, l*=2]
         [$\lnot \lock{22}$, l*=2]
       ]
       [$\lnot \red{21}$, l*=3
         [$\green{21}$]
       ]
    ]
    [$\lnot \safe(\rt{2131})$, before computing xy={s/.average={s}{siblings}}
      [$\lnot \unoccupied(\rt{2131})$
        [$\occupied(\track{1})$
          [$\lnot \tc{1}$]
        ]
      ]
    ]
    [$\lnot \safe(\rt{2133})$
      [$~~\lnot \routelocked(\rt{2133})$
        [$\lnot \pointsok(\rt{2133})$
          [$\lnot \ptleft{}(\pt{21})$
            [$\lnot \ptleft{21}$]
          ]
        ]
      ]
    ]
  ] 
\end{forest}
$$
\caption{Explanation tree}{}
\label{tree}
\end{figure}

$\expl$ is again applied to each of the formulæ to create new branches and definitions are again applied to each new explanation literal, and so on until no more explanations are obtained. This gives the explanation tree in Figure \ref{tree}.

The tree is cut off at the second occurrence of $\green{21}$, as there already is an explanation for that literal being true.

One explanation of the failure is that signal 21 is showing a ``proceed'' aspect. Further explanations show that the green light in the signal is turned on and the red light turned off, which is expected. Another explanation is that the route from signal 21 to signal 31 is not safe. Further explanations show that this is due to the route not being unoccupied, to track 1 being occupied and finally to input signal $\tc{1}$ not being true. The interlocking logic for signal 21 (definition of $\green{21}$) indeed does not depend on $\tc{1}$ so this is the reason for the failure.

One branch remains. Signal 21 is also the starting point of the unsafe train route to signal 33. That, however, turns out to be because points 21 are not set to the left, which is expected as the route takes the right path through the points. We can see that the interlocking does check that condition properly.

To see some details, consider $\lnot\unoccupied(\rt{2131})$. The definition of $\unoccupied$ gives $\lnot\forall u.(\partof(u,\rt{2131})\limp \lnot \occupied(u))$. Applying $\expl$ gives
$$
\begin{array}{l}
\expl(\lnot\forall u.(\partof(u,\rt{2131})\limp \lnot \occupied(u))) \\
= \expl(\forall u.(\partof(u,\rt{2131})\limp \lnot \occupied(u))) \\
= \expl(\rewrite(\forall u.(\partof(u,\rt{2131})\limp \lnot \occupied(u)))) \\
= \expl((\partof(\track{1},\rt{2131})\limp \lnot \occupied(\track{1}))~\land \\
\phantom{= \expl(} (\partof(\track{2},\rt{2131})\limp \lnot \occupied(\track{2}))\land \dots)
 \end{array}
 $$
All conjuncts of the last argument to $\expl$ are true except the first one which is false. According to Table \ref{expltable}, only the explanation of that conjunct will contribute to the explanation of the whole formula, so we continue
$$
\begin{array}{l}
= \expl(\partof(\track{1},\rt{2131})\limp \lnot \occupied(\track{1})) \\
= \expl(\rewrite(\partof(\track{1},\rt{2131})\limp \lnot \occupied(\track{1}))) \\
= \expl(\lnot(\partof(\track{1},\rt{2131})\land \lnot\lnot \occupied(\track{1}))) \\
= \expl(\partof(\track{1},\rt{2131})\land \lnot\lnot \occupied(\track{1})) \\
 \end{array}
 $$
Now both conjuncts of the last argument to $\expl$ are true so, again according to Table \ref{expltable}, the one with the highest interest measure is chosen. As we mentioned above, $\occupied$ is of higher interest than $\partof$, so the final explanation becomes the explanation of $\lnot\lnot\occupied(\track{1})$, which is $\{\occupied(\track{1})\}$.

\section{Discussion}
\label{discussion}

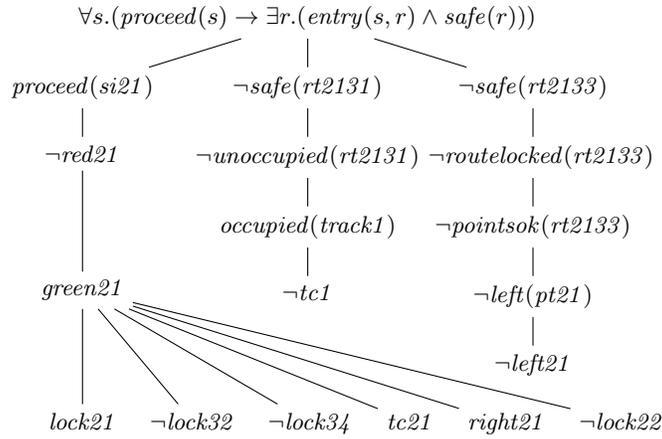
\begin{figure}[!htp] 
$$
\begin{forest}
   [${\forall s.(\proceed(s) \limp  \exists r.(\entry(s, r) \land \safe(r)))}$
     [$\proceed(\si{21})$
       [$\lnot \red{21}$
         [$\green{21}$, l*=2, calign=first
           [$\lock{21}$, l*=2]
           [$\lnot \lock{32}$, l*=2]
           [$\lnot \lock{34}$, l*=2]
           [$\tc{21}$, l*=2]
           [$\ptright{21}$, l*=2]
           [$\lnot \lock{22}$, l*=2]
         ]
       ]
    ]
    [$\lnot \safe(\rt{2131})$, before computing xy={s/.average={s}{siblings}}
      [$\lnot \unoccupied(\rt{2131})$
        [$\occupied(\track{1})$
          [$\lnot \tc{1}$]
        ]
      ]
    ]
    [$\lnot \safe(\rt{2133})$
      [$~~\lnot \routelocked(\rt{2133})$
        [$\lnot \pointsok(\rt{2133})$
          [$\lnot \ptleft{}(\pt{21})$
            [$\lnot \ptleft{21}$]
          ]
        ]
      ]
    ]
  ] 
\end{forest}
$$
\caption{Alternative explanation tree}{}
\label{tree2}
\end{figure}

While pinpointing the relevant reason for the failure, the explanation tree of Figure \ref{tree} also includes some redundant information that does not contribute to the understanding of the failure. For example, the subtrees beginning with $\lnot \red{21}$ and $\lnot \safe(\rt{2133})$ do not point to the actual cause of failure. The subtree beginning with $\lnot \red{21}$ could be seen as redundant as it is cut off at $\green{21}$ which has occurred earlier in the tree. However, even with such pruning if the branch with $\lnot \red{21}$ had been developed before the branch with $\green{21}$, we would get the tree of Figure \ref{tree2}. This tree arguably does not give a better explanation as it has the ``detour'' through $\lnot \red{21}$ to get to the explanations of $\green{21}$. 

The explanation trees have three distinct branches of which only one actually gives the desired explanation. It is not clear how to algorithmically pinpoint that branch as that would require an actual understanding of the purpose of the system being verified. The explanation that signal 21 shows a proceed aspect \textit{is} valid, as forcing the signal to red at all times would make the verification succeed. However, that would also make the interlocking system unfit for its intended purpose. The remaining explanation ($\lnot\ptleft{21}$) could in principle have been caused by an incorrect check of the points position for train route $\rt{2133}$.

The false-false case of Table \ref{expltable} could appear to lose information when the interest levels of the two explanations are the same as only one of the explanations of the two conjuncts, $P$ and $Q$, is used. This could matter if the falsity of the conjuncts is independently caused by faults in the system being verified. For example, if the quantified subformula of the example verification condition, $\proceed(s) \limp  \exists r.(\entry(s, r) \land \safe(r)))$, was false for two or more different signals, $s$, only one signal would figure in the explanation.

However, information would not actually be lost in a practical verification setting. Generally all faults in a system will not manifest in a single verification counterexample. Once one fault has been identified, to find another one the verification must be retried with assumptions added that exclude the situation leading to the identified fault, generally giving a different counterexample\footnote{This methodology is further described in \cite{GTO}.}. 

In the case of $P\land Q$ where $P$ and $Q$ are false because of independent faults and the explanation of $P$ is chosen, excluding that fault situation in a new verification attempt would make $P$ true and then the explanation would be based on $Q$. For the same reason it doesn't matter for completeness that when the interest levels of the two explanations are different, the least interesting one is used.

In the railway example, changing the false-false case with equal interests to use both explanations would not add any more information to the explanation tree. It would, however, create an additional short branch in the tree which would be redundant in the same way as the $\lnot\red{21} - \green{21}$ branch.

\section{Related work}
\label{relatedwork}

Broadly speaking, there are two main approaches to explaining the causes of counterexamples -- with and without the presence of a domain model. In the approach without a domain model, the explanation focuses on what atoms in the model of the system to be verified have values that ``cause'' a verification condition to fail. In approaches with a domain model, the explanation is given in terms of that model. Our approach is primarily an example of the latter.

Most work on explanation of interpretations in terms from a theory/model have been done in the context of the Alloy Analyzer \cite{DBLP:journals/cacm/Jackson19}. This includes \textit{model exploration} which is technically very similar to counterexample explanation. The purpose of explanation in model exploration is to identify the formulæ that ``cause'' a particular aspect of a satisfying interpretation for the purpose of helping a user to understand or debug a model of some software or system.

The second phase of our method finds explanations of literals using definitions. This has the drawback that literals with predicates that are not explicitly defined can't be explained. Nelson et al. \cite{nelson2017power} give an algorithm to find \textit{provenances} -- explanations of literals using arbitrary formulæ –– not just definitions. A provenance is a finite set of formulæ which explain why a literal is true in the sense that the literal is a logical consequence of the provenance (and the model), while the provenance formulæ are independent of the literal in the sense that they remain true even if the truth value of the literal would be reversed.
 
We will compare their approach to ours, using the interpretation of definitional formulæ as equivalences. As the description in \cite{nelson2017power} uses the relational logic of the Alloy Analyzer\footnote{As is the case with our logic, quantifiers in the logic of Alloy are bounded.} \cite{DBLP:journals/cacm/Jackson19}, we have reformulated it in appendix \ref{provenances} in the context of predicate logic. Also, as the description in \cite{nelson2017power} is somewhat hard to follow, we have modified it slightly for simplicity and clarity. Thus our formulation of the provenance algorithm can also be of interest in its own right.  

In the example, the explanation of the starting formula was the set $$\{\proceed(\si{21}), \lnot \safe(\rt{2131}), \lnot \safe(\rt{2133})\}$$  For these three literals, the provenance algorithm gives the same result as our use of definitions. However, further down the tree, using the provenance algorithm on $\lnot \unoccupied(\rt{2131})$ gives two possible explanations. One is the same we get using definitions, $\lnot \forall u.(\partof(u,\rt{2131})\limp\lnot \occupied(u))$, the other is the set $$\{\lnot \safe(\rt{2131}), \routelocked(\rt{2131})\land \pointsok(\rt{2131}), \lnot \inconflict(\rt{2131})\}$$  $\lnot \safe(\rt{2131})$ is redundant, having already been encountered in the tree. The other two formulæ do not point to facts relevant to the failure. The reason is that the provenance algorithm lacks the sense of direction that definitions provide. In this example, by pruning explanations including literals already encountered (such as $\lnot \safe(\rt{2131})$), we actually do obtain the same tree as in Figure \ref{tree}.

In some cases the lack of a sense of direction could be advantageous. For example, using the axiom $\lnot \ptleft{21} \lor \lnot \ptright{21}$, the provenance algorithm explains $\lnot \ptleft{21}$  by $\ptright{21}$ being true. In conclusion, while the provenance algorithm could give some additional explanations, its lack of direction also causes difficulties.

Dougherty et.al. \cite{explorationsurvey,saghafi2015exploring,saghafi2014razor} have studied how finding satisfying interpretations which are, in a sense, minimal can help provide explanations. Zheng et.al. \cite{zheng2021flack} describes a fault localisation method for Alloy models using SAT solving to find interpretations that exhibit a fault as well as those that do not. Failing and non-failing interpretations are compared and the differences are used to pinpoint parts of the model likely to contribute to the failure. Wang et.al. \cite{khan2021alloyfl,wang2020fault} use a test-driven approach, which is similar in that interpretations are compared, but where they are generated by mutating failing properties. Satisfying interpretations using different tests are compared to find parts of the model that likely contribute to the failure.

The work of Siegel et.al. \cite{siegel2009untwist} does not use a domain theory but otherwise has some similarities to ours, pinpointing the parts of verification conditions that contribute to the failure (i.e. to the condition being false).

\section{Conclusions and further work}
\label{conclusions}

We have devised and implemented a method for giving high-level explanations of failures of verification proofs carried out by SAT solvers. The main contribution is combining the use of a domain theory to define high-level concepts with a concept of interest to extract the information most likely to give a useful explanation. The method has been demonstrated using a small but realistic example. Some shortcomings have been discussed which could be addressed by further work. Also possible extensions using the related work of Nelson et al. \cite{nelson2017power} have been discussed. It would also be of interest to implement the method for the Alloy Analyzer to gain experience with the large amount of models developed for Alloy.

Natural extensions of our approach would be to use temporal logic and unbounded quantification (infinite domains). Extension to linear time temporal logic (LTL) with past-time operators \cite{lichtenstein1985glory} seems straightforward. Instead of a single interpretation, we would have a sequence of interpretations (a trace). The explanation algorithm would first use the final interpretation of the sequence and move back one step in the sequence when encountering the previous-moment operator, $\circleddash$. The \textit{since} connective, $S$, could be handled by replacing every (sub)formula $P\ S\ Q$ with a new unique predicate $psq$ and for all such predicates adding the axiom $psq \lequiv P \lor Q \land \circleddash psq$. As past-time and standard (future-time) LTL have the same expressibility over finite traces \cite{lichtenstein1985glory}, formulæ of standard LTL can be handled by translating into past-time LTL.

Handling unbounded quantification would allow explanations of failing interpretations generated by SMT solvers \cite{SMT} and could also be advantageous when quantifying over finite domains so large that the basic method gives unwieldy results. How to accomplish this is less clear, as the first phase of our method relies on expanding universal quantifiers to a conjunction of instances of the quantified formula. In the case where the quantified formula is false, according to Table \ref{expltable} the explanation of any single formula instantiated with a witness to the falsity will do. In the case where the quantified formula is true, then -- again according to Table \ref{expltable} -- the explanation should be made up of the explanations of all the infinite number of instances, which is not possible. 

In the case that true instances have uniform explanations (identical except for the instantiation of the quantified variable), a single explanation including the quantified variable could be given. That approach could also be used in the case of bounded quantification, in particular when the domain is large. The question then, is how to determine that all instances have uniform explanations? On the other hand, the Alloy Analyzer has been very successful using only finite domains, so maybe an extension to unbounded quantification is not so important.

\appendix
\section{Provenances}
\label{provenances}

This appendix gives the provenance algorithm of \cite{nelson2017power} reformulated for clarity and using predicate logic rather than relational logic.

Given a theory (set of formulæ) $\mathcal{T}$, a satisfying interpretation $\mathbb{M}$ of $\mathcal{T}$ with a finite universe $\mathcal{U}$, and a  literal $\mathcal{L}$ true under  $\mathbb{M}$, we want to explain why $\mathcal{L}$ is true. A \textit{provenance} for $\mathcal{L}$ is a finite (possibly empty) set of true formulæ $P_i$ which explains why $\mathcal{L}$ is true in the sense that $\mathcal{L}$ is a logical consequence of the $P_i$ and $\mathcal{T}$ while at the same time, each $P_i$ will remain true even if the truth value of $\mathcal{L}$ is reversed. (I.e. the truth of the $P_i$ do not depend on the truth value of $\mathcal{L}$.) 

\begin{definition}[$\mathcal{L}$-alternate]
Given an interpretation $\mathbb{M}$ and a true literal $\mathcal{\mathcal{L}}$, the \textit{$\mathcal{L}$-alternate}, $\mathbb{M}^\mathcal{L}$, is the same interpretation as $\mathbb{M}$ except that $\mathcal{L}$ is false, i.e. the atom of $\mathcal{L}$ has the different truth value in $\mathbb{M}^\mathcal{L}$.
\end{definition}

\begin{definition}[Provenance]
A \textit{provenance} for a literal $\mathcal{L}$ (true in $\mathbb{M}$), with respect to a set of formulæ $\mathcal{T}$ (all true in $\mathbb{M}$),  is a set of formulæ $\{P_1,\dots,P_n\}$, for $n\geq 0$, such that each $P_i$ is true in both $\mathbb{M}$ and $\mathbb{M}^\mathcal{L}$, and $\mathcal{T}\cup \{P_1,\dots,P_n\} \vDash \mathcal{L}$.
\end{definition}

Provenances are obtained by the ``explanation function'' $\mathcal{Y}$, parameterised by $\mathbb{M}$ and $\mathcal{L}$, so that $\mathcal{Y}(\mathcal{T})$ gives the set of all provenances of $\mathcal{L}$.  $\mathcal{Y}$ is defined on formulæ true in $\mathbb{M}$ but false in $\mathbb{M}^\mathcal{L}$. Strictly speaking we must also have $\mathbb{M}^\mathcal{L} \nvDash \mathcal{T}$, but if we did not  $\mathcal{L}$ would be independent of $\mathcal{T}$ and then there would be no provenance anyway.

The auxiliary function $\mathcal{Y}_{\lnot}$ handles the dual case, being defined on formulæ false in $\mathbb{M}$ but true in $\mathbb{M}^\mathcal{L}$. The auxiliary function $\rewrite$ (parameterised by $\mathcal{U}$) is as defined  in section \ref{finding}.

\begin{definition}[$\mathcal{Y}$]
$$
\begin{array}{ll}
\mathcal{Y}(A) = \{\emptyset\}, \text{if $A$ is an atom}\hfill \\
\mathcal{Y}(P_1 \land \dots \land P_n) = \bigcup \{\mathcal{Y}(P_i)~|~1\leq i\leq n, \mathbb{M}^\mathcal{L}\nvDash P_i \} \\
\mathcal{Y}(\lnot P) = \mathcal{Y}_{\lnot}(P) \\
\mathcal{Y}(P) = \mathcal{Y}(\rewrite(P)), \text{in all other cases} \\
\\
\mathcal{Y}_{\lnot}(A) = \{\emptyset\}, \text{if $A$ is an atom} \\
\mathcal{Y}_{\lnot}(P_1 \land \dots \land P_n \land Q_1 \land \dots \land Q_m) = \\
 ~~\{\{P_1,\dots, P_n\}\cup P~|~P\in (\mathcal{Y}_{\lnot}(Q_1) \stackrel{_{\cup}}{\times}\dots \stackrel{_{\cup}}{\times} \mathcal{Y}_{\lnot}(Q_m))\},  \\
 ~~~\text{where }\mathbb{M}\vDash P_i, 1\leq i\leq n\text{ and }\mathbb{M}\not\vDash Q_j, 1\leq j\leq m \\
\mathcal{Y}_{\lnot}(\lnot P) = \mathcal{Y}(P) \\
\mathcal{Y}_{\lnot}(P) = \mathcal{Y}_{\lnot}(\rewrite(P)), \text{in all other cases} \\
\\
A\stackrel{_{\cup}}{\times}B = \{a\cup b~|~a\in A, b\in B\} \\
\end{array}
$$
\end{definition}

\noindent Note that the requirement that the argument of $\mathcal{Y}$ is true in $\mathbb{M}$ but false in $\mathbb{M}^\mathcal{L}$ means that in the atom case, the argument can only be $\mathcal{L}$. Similarly for $\mathcal{Y}_{\lnot}$ where the negation of an atomic argument must be $\mathcal{L}$. Also, in the case for $\land$ in $\mathcal{Y}_{\lnot}$, we could have $n=0$ but must have $m>0$ as the argument of $\mathcal{Y}_{\lnot}$ is always false.

\begin{example}
(Adapted from \cite{nelson2017power}.) We give a simple theory for undirected graphs where nodes are coloured either blue or red and are coloured blue iff they have at most one neighbour.  The universe is the set of nodes. The predicate $\neighbours$ relates nodes which are connected by an edge.  The set $\mathcal{T}$ consists of the formulæ
$$
\begin{array}{ll}
\forall x.\forall y.(\neighbours(x,y) \limp \neighbours(y,x)) \\
\forall n.(\red{}(n) \lequiv \lnot \blue(n)) \\
\forall n.(\blue(n) \lequiv \lnot \exists x.\exists y.(\neighbours(n,x) \land \neighbours(n,y)\land x\neq y)) \\
\end{array}
$$
\end{example}

Let $\mathbb{M} = \{\neighbours(n_0,n_1), \neighbours(n_1,n_0), \blue(n_0), \blue(n_1)\}$ with $\mathcal{U} = \{n_0, n_1\}$. Now,  we ask why $\mathcal{L} = \blue(n_0)$ is true.$$\mathcal{Y}(\mathcal{T}) = \{\{\lnot \red{}(n_0)\}, \{\lnot \exists x.\exists y.(\neighbours(n_0,x) \land \neighbours(n_0,y)\land x\neq y))\}$$

\noindent That is, there are two distinct reasons why $n_0$ is blue. 1) it is not red and 2) it has at most one neighbour.

\newpage

\bibliographystyle{splncs04}
\bibliography{explanation}

\end{document}